# FIRST-PASSAGE TIME: A CONCEPTION LEADING TO SUPERSTATISTICS.
## I. SUPERSTATISTICS WITH DISCRETE DISTRIBUTIONS

V. V. Ryazanov*

*Institute for Nuclear Research, pr. Nauki, 47, 03068, Kiev, Ukraine*

**Abstract**

To describe the nonequilibrium states of a system we introduce a new thermodynamic parameter - the lifetime (the first passage time) of a system. The statistical distributions that can be obtained out of the mesoscopic description characterizing the behaviour of a system by specifying the stochastic processes are written. Superstatistics, introduced in [1] as fluctuating quantities of intensive thermodynamical parameters, are obtained from statistical distribution with lifetime (random time to system degeneracy) as thermodynamical parameter (and also generalization of superstatistics). Necessary for this realization condition with expression for average lifetime of stationary statistical system obtained from stochastical storage model [32] is consisting. The obtained distribution passes in Gibbs distribution depending on a measure of deviation from equilibrium related to fluxes and dissipativity in the system.

----------------------

*Corresponding author. E-mail address: vryazan@kinr.kiev.ua

# 1. Introduction

In works [1, 2] the generalization of Boltzmann factor $exp\{-\beta_0 E\}$ was introduced in the following form:

$$B(E) = \int_0^\infty d\beta` f(\beta`) exp\{-\beta`E\}. \qquad (1)$$

It is supposed there that the intensive parameter (return temperature $\beta$, chemical potential, etc.) fluctuates. These fluctuations evolve on a long time scale. Locally, in some spatial area (cell) where $\beta$ is approximately constant, the system is described by usual Boltzmann-Gibbs statistics with ordinary Boltzmann factor $exp\{-\beta E\}$ where $E$ is the energy a microstate associated with each cell. On a long time scale it is necessary to take into account fluctuations of $\beta$. Superposition of two statistics (that of $\beta$ and that of $exp\{-\beta E\}$) which therefore and refers to as "superstatistics" is derived. This formalism is successfully applied to the description of fully developed hydrodynamic turbulence, defect motion in convection states, the statistics of cosmic rays and other metastable and nonequilibrium phenomena. The special case of these superstatistics, at function $f$, reduces to gamma-distribution, appears in the nonextensive statistical mechanics [3, 4], describing a number of the physical phenomena which are not satisfactory described by Boltzmann-Gibbs statistics (for example, long-range many body systems, systems with memory, many phenomena in nuclear physics, astrophysics, geophysics, ecology and other complex systems). In the present work the superstatistics as (1) (together with its generalization) is obtained starting from nonequilibrium thermodynamics which as a thermodynamic variable contains a first-passage time, lifetime of statistical system [5, 6, 7, 8], Section 2.

It is mean explaining what is understood under the notion of a first-passage time or lifetime. Arbitrary systems are formed from elements, particles, or elementary objects. These elements enter any system and leave it. It is possible to formulate the notion of a lifetime with mathematical strictness; one thus understands the time period at which the elements are present in a system. For example, if we consider a small volume cell of a gas, its lifetime is the period till all molecules occasionally leave this cell. Are examples borrowed from chemical kinetics are times at which the number of molecules of certain kind becomes zero, time of the dissociation of a two-atom molecule, the age of a (biological) system, residence times of grains in sand-pile models etc.

Stratonovich in [9] used the term "lifetime" as terminus techniques with respect to the number of phenomena considered. Following synonyms are encountered by us: the first passage time (for some given level), escape time, busy period (in the queuing theory) etc. First published paper on the subject is [10] where the Pontryagin equations for the lifetime distribution are obtained. In 1940 the Kramers work [11] appeared which dealt with the escape time from the potential well. These questions are discussed in books by van Kampen [12], Gardiner [13] and other authors [14]. The lifetime plays part in the theory of phase transitions, chemical reactions, in investigating dynamics of the complex biomolecules, calculating the coefficient of the surface diffusion in semiconductors, in nuclei, elementary particles, spin glasses, spectroscopy, trap systems, in the theory of metastable states etc.

The lifetime is the time period till the moment of the first (random) reattachment of a certain level (for example, zero level) by a random process $y(t)$ for the macroscopic parameter (2) (for example, energy or particle number). The lifetime is the slave process in the terminology of the random processes theory; that of $y(t)$ (*Fig.1*) determines the behaviour of lifetime. The lifetime depends on the energy of a system, its size, fluxes of energy. Therefore a system exchanges with thermostat the energy, the particle number, but not the "lifetime". The lifetime $\Gamma$ is a macroparameter characterizing the system and its interaction with the environment. It is an observable, well-defined and physically well understandable quantity reflecting important system peculiarities.

We don't consider the current time, but rather the lifetime of a specific system, its age determined by the evolution of the system. Vice versa: the properties of a system depend on its age (such as for a human, evidently, the age is of prominent importance). In the thermodynamics and statistical physics it is supposed that statistical systems tend to the state of equilibrium. We consider open systems for which the equilibrium state is vague concept. But each and every physical system possesses finite lifetime, the latter being thus a universal physical characeristics. We don't consider the idealised systems of infinite size, which is we don't perform the thermodynamic transition, as in [15,16]. Many systems degenerate without reaching equilibrium. Therefore lifetime seems to be more representative value than the equilibrium reaching time. In [4,5,6] the conception of the nonequilibrium thermodynamics with lifetime as a thermodynamic parameter is derived.

In [17] it is pointed that a nonequilibrium distribution is characterized by an additional macroparameter in the description. One can show [18] that in the method of nonequilibrium statistical operator (*NSO*) [19, 20] this parameter is the time $t-t_0$, from the birth of a system, that is this is the first-passage time of the zero level if we inverse time, the age of the system (in Zubarev work [19] the *NSO* is obtained by averaging over the initial time; in [21] it is noted that the $\varepsilon$ value from *NSO* in [19, 20] is inversely proportional to the lifetime). In works of Kirkwood [22] it was noted that the state of a system at the current time depends on the whole evolution history of the nonequilibrium processes in it, thus, on its age as well. Therefore we choose the random lifetime as an additional thermodynamical parameter. In [18] an analogy is outlined between the approaches of Prigogine [23] and that of Zubarev [19, 20] - the latter approach considers the dynamics of a system and correlations from the whole past of a system.

When considering the finite size systems the finiteness of lifetime seems to be essential. Normally functioning system appears to be in a steady nonequilibrium state characterized by a given deviation from equilibrium and the entropy production. Each state of a system has its own lifetime related to fluxes magnitudes and sources strengths and therefore its deviation from equilibrium. The influences experienced by a system interacting with environment lead to the deviation from the normal stable state and to the dissipative effects, they equally change the degree of deviation from the stationary state, entropy and lifetime. Raising lifetime $\Gamma$ to the realm of the macroscopic observables reflects possible complication of the structure of the phase space that can contain domains of different behaviour (for example, attractors with anomalously big lifetimes etc). From the random processes theory it follows that the existence and finiteness of a lifetime is provided by the existence of stationary states, physically meaning the existence of stationary structures.

One considers an open system, and dynamical characteristics under the interaction with the surrounding turn to be random; thus the macrodescription only is possible. The structure of phase space generally has complicated (fractal) shape, conditioned by the noncontinuous character of the underlying stochastic process. The lifetime is macroscopic value; one does not need explicit dependence of the lifetime on the phase space variables. Likewise in the approach based on the maximum entropy principle [24] one doesn't need the knowledge of the details of the microscopic behaviour. Therefore one introduces lifetime as a phenomenological macroscopic value. It is not a dynamical value (although it depends on the phase space coordinates), but it is a statistical value described by well-known equations for the density of its distribution function (for example, the Pontryagin equations [10]).

The open systems are characterized by energy and entropy redistribution, nonhomogeneous character of the time flow. The events in the system history are conditioned by the entropy and information fluxes from inside and from outside. For example, we determine from the breath whether a human is alive. In the theory of random processes there is a key notion where from it is possible to derive the whole theory, that is the notion of the stop moments, and the lifetime is a particular kind of them. The events bearing the information on a system create the time flux in it; the moments of birth and death stay apart. Since we consider open systems which exchange the dynamical quantities with surrounding the introduction of the lifetime as a principal random value is believed to be an important and necessary element of correct description of nonequilibrium phenomena.

It is worth noting that the notion of lifetime has many interpretations, and its physical applications are also various. For example, the relaxation times can be considered as lifetimes of the excited states. An analogy can be traced between the approach of the present work and the description of nonergodicity in the spin glasses [25] (although the "lifetime" term was not used there), which included the memory effects (dependence on the previous history of a system), dependence on the initial time moment, and on the times of transition between different valleys on the energy profile. This problem is related to the response functions in systems with excited degrees of freedom [26]. In our work we presented the lifetime, likewise it was done in the pioneering work [10], from the outset in a strict mathematical form (2), similarly to the same value introduced in the theory of random processes. One can admit various physical interpretations of the general mathematical description. One of these possibilities applied in this work to the lifetime of the system in the energetic valleys, leads to the obtained class of statistical distributions containing the product of Gibbs-like factor and that of superstatistics.

In [27] it is noted that the *q*-exponential pdf's represent a special case, which in turn includes the Boltzmann-Gibbs pdf's as an even more special one, corresponding to the standard thermal equilibrium. It seems natural to believe that realistic physical systems combine both the properties of Gibbs systems and those of nonextensive statistical mechanics. Therefore the investigation of suggested distribution seems to be prospective, which is confirmed by the existence of such kind of distributions for a number of various phenomena. It schould be noticed that in this formalism the distribution is obtained in the first place.

The paper is organized as follows. In Section 2 we consider the statistical distribution with lifetime [4-6]. In Section 3 the examples of unperturbed equilibrium and stationary distributions for lifetime are considered; they enter the expression for the generalized structure $\omega(E,\Gamma)$ (number of states with given macroparameters). In Section 4 we make assumptions as to the shape of $\omega(E,\Gamma)$. As in self-organized criticality (*SOC*) [28] it is essential that the system have many metastable states. *SOC* a finite-size-scaling hypothesis is successfully tested. The statistical distribution is obtained which represents a product of a Boltzman-Gibbs factor, and the superstatistic factor. The weight of each factor depends on the deviation from equilibrium and on the fluxes magnitude. The examples of application of the lifetime concept and use of expressions with a lifetime for various concrete physical problems are resulted in Section 5. We finish with a discussion and summary in Section 6.

## 2. System lifetime and lifetime distribution

In the present work we consider open nonequilibrium systems, stationary nonequilibrium systems, certain point metastable states. Work on spin glasses and other aging systems, where a "waiting time" plays an important role, points in this direction.

Characterizing the nonequilibrium state by means of an additional parameter related to the deviation of a system from the equilibrium (field of gravity, electric field for dielectrics etc) was used in [17]. In the present paper we suggest a new choice of such an additional parameter as the lifetime of a physical system which is defined as a first-passage time till the random process *y(t)* describing the behaviour of the macroscopic parameter of a system (energy, for example) reaches its zero value (*Fig.1*). The lifetime $\Gamma_x$ (or $\Gamma$) is thus a random process which is subordinate (in terms of the definitions of the theory of random processes [29]) with respect to the master process $y(t)$,

$$\Gamma_x = \inf\{t: y(t)=0\}, \ y(0)=x>0 \ . \tag{2}$$

This definition of the lifetime is taken from the theory of random processes where it is widely used in the theory of queues, stochastic theory of storage [48]. There it is possible to find numerous examples of his application. The characteristics of $\Gamma$ depend on those of $y(t)$. We use below as $y(t)$ the value of full energy of the system $E$, assuming her nonnegativity though it is possible to consider as $y(t)$ and (or) the number of particles in the system. For example, for an rarefied gas flowing through a pipe $y(t)$ is the number of particles of gas in the pipe, and $\Gamma$ is time during which in a pipe there are particles of gas.

It is important to make clear a physical interpretation to the definition (2). So, the lifetime is related to the period of stable existence of a system, its time of dwelling within the homeokinetic plateau whose distribution was related in [30] to the entropic and information parameters of a system, its response to internal and external influences, its stability and adaptation facilities. The states of a system within the homeokinetic plateau are characterized by the mutual compensation of the entropic effects related to the energy dissipation and by the effects of negative entropy determined by the existence of the negative feedbacks. When the system exits out of the limits of the plateau unstable structures and sharp qualitative changes in the behaviour of the system arise. We assume that the existence and the magnitude of finite lifetime are related to the deviation of the system from the equilibrium.

In the works on statistical physics similar definitions of thermodynamic values are encountered. For example, in [31]: "Any function $B(z)$ of dynamic variables ($z=q_1,...,q_N, p_1,...,p_N$), having macroscopical character, by definition is random internal thermodynamic parameter". In [17] all values dependent on $z$ concern are treated in this way. The fact that a lifetime (2) $\Gamma(z)$ is the function of $z$, is obvious from the equations for distribution of a lifetime in Markov model [10], which are related to the equation the $z$-dependent density of distribution. Pontryagin equations for density of probability of a lifetime are conjugated to the equations such as the Fokker-Planck equations for density of probability of energy (which depends from $z$). Evidently it is visible from a situation when one considers as system some volume with gas in which all molecules are located on its borders with the velocities directed outwards. Then a lifetime is finite. For other configuration of coordinates and pulses a lifetime will be another. The lifetime values are influenced by the attractors, metastable states, phase transitions and other physical peculiarities of a system which depend on $z$. The same is clear from the consideration of the examples of: a) stochastic storage model in which the lifetime is proportional to the statistical sum [32]; b) the mean lifetime of a neutron subsystem after integration over velocities and coordinates [18]; c) the sedimentation times for aerosols [33]. Besides that the lifetime shows the dependence on the energy $E$ as well as particle number $N$ which are likewise $z$-dependent. Therefore we consider the lifetime to be a thermodynamic parameter.

To validate the choice of lifetime as thermodynamic value one can reassert to the method of Zubarev *NSO* [19, 20] which is interpreted in [18] as averaging of quasiequilibrium (relevant) the statistical operator on distribution of lifetime of the system $p_q(y)$. Then at $p_q(y)=\varepsilon\exp\{-\varepsilon y\}$ (in [18] a more general choice, merely arbitrary one, of the function $p_q(y)$ is suggested)

$$\ln\rho(t) = \int_0^\infty p_q(y)\ln\rho_q(t-y,-y)dy = \ln\rho_q(t,0) - \int_0^\infty \exp\{-\varepsilon y\}(d\ln\rho_q(t-y,-y)/dy)dy \ , \tag{3}$$

where $\rho$ is nonequilibrium statistical operator, $\rho_q$ is relevant statistical operator [19, 20]. The entropy production operator [19, 20] is equal $\sigma(t-u, -u) = d\ln\rho_q(t-u, -u)/du$. If $\sigma(t-u, -u) \approx \sigma(t)$ or $\sigma(t-u,-u) \approx <\sigma(t)>$ has weak $u$-dependence then equality (3) acquires a form $ln\rho(t)=ln\rho_q(t,0)+<\sigma(t)>\varepsilon^{-1}=ln\rho_q(t,0)+<\sigma(t)><\Gamma>$, as in interpretation [18] $\varepsilon^{-1}=<\Gamma>=<t-t_0>$ is average lifetime. We shall note, that such distribution is received in [34]. Then a relation (3) in nonequilibrium thermodynamics with lifetime as thermodynamical parameter is replaced with expressions (5) - (6), or, in more general form, $ln\rho(t)=ln\rho_q(t,0)+ln[Z(\beta)/Z(\beta,\gamma)]-\gamma\Gamma$, where $\rho=exp\{-\beta E-\gamma\Gamma\}/Z(\beta,\gamma)$, $\rho_q=exp\{-\beta E\}/Z(\beta)$, $Z(\beta)=\int exp\{-\beta E\}dz$ (in (5) instead of $<\Gamma><\sigma(t)>$ stand the random value $\Gamma$ and Lagrange multiplier $\gamma$ in: $\gamma\Gamma+lnZ(\beta,\gamma)/Z(\beta)$). That is, averaging on distribution of lifetime in Zubarev *NSO* is replaced with use of a random variable of lifetime, - there lays essential difference of distribution (5) - (6) from (3) and from results of works [19, 20, 34].

The value $\Gamma_x$ in (2) it is time before degeneration, destruction of the system. From the theory of the random processes it follows that existence and finiteness of the lifetime $\Gamma$ value is related to the presence of stationary states, which physically corresponds to the existence of stationary structures. Complex functional and hierarchical relations of real systems correspond to the analogous between lifetimes determining evolutionary processes as sequence of transitions between various classes of the system states. In contrast to the traditional representation about time as about something changeable, lifetime acts as result of existence of stable stationary structures. It depends on external influences on system and from internal interactions in it.

In termostatics it is supposed that any isolated thermodynamic system eventually reaches an equilibrium state and never spontaneously leaves it. However the assumption of a possibility of fluctuations in system (i.e. random deviations of internal parameters from their equilibrium values) contradicts main principles termostatics. We shall replace the assumption of a system remaining in the basic (equilibrium) state for an infinite long time at its isolation from influence of environment with a more physical assumption (corresponding also to the statistical description) about an opportunity of a system leaving this state and destruction, degeneration of the system under influence of internal fluctuations.

The distribution for lifetime $\Gamma$ (2) generally depends on macroscopical value $y(t)$ of the master process that governed by $\Gamma$, the master random process describing some relevant macroparameter (energy or particle number) (*Fig.1*). Let us suppose that the process $y(t)=E$ is the energy of a system (equivalently one could choose the particle number, pulse etc). Gibbs equilibrium distribution for microcanonic probability density in phase space $z$ corresponds to a condition of equiprobability of all possible microstates compatible with the given value of a macrovariable. We shall assume, that transition of system in a nonequilibrium condition breaks equality of probabilities, characteristically for an equilibrium case. One introduces additional observable macroparameter, thus extending the phase space (containing additional degenerate, absorbing states). We shall assume validity of a principle of equal probabilities for the extended phase space divided into cells with constant values $(E, \Gamma)$ (instead of phase cells with constant values $E$). The standard procedure (e.g. [35]) allows one to write down the relation between the distribution density $p(E,\Gamma)=p_{E\Gamma}(x,y)$ and microscopic (coarse-grained) density $\rho(z;E,\Gamma)$

$$p(E, \Gamma)=\int \delta(E-E(z))\delta(\Gamma-\Gamma(z))\rho(z; E, \Gamma)dz=\rho(z;E, \Gamma)\omega(E, \Gamma). \qquad (4)$$

The structure factor $\omega(E)$ is thus replaced by $\omega(E,\Gamma)$ - the volume of the hyperspace containing given values of $E$ and $\Gamma$. If $\mu(E,\Gamma)$ is the number of states in the phase space which have the values of $E$ and $\Gamma$ less than given numbers, then $\omega(E,\Gamma)=d^2\mu(E,\Gamma)/dEd\Gamma$. It is evident that $\int\omega(E,\Gamma=y)dy=\omega(E)$. The number of phase points between $E,E+dE$; $\Gamma,\Gamma+d\Gamma$ equals $\omega(E,\Gamma)dEd\Gamma$. We make use now of the principle of equiprobability applied to the extended cells $(E,\Gamma)$.

Using a maximum-entropy principle [24], it is possible to write down expression for microscopic (but coarse-grained) density of probability in the extended phase space

$$\rho(z; E, \Gamma)=exp\{-\beta E-\gamma\Gamma\}/Z(\beta,\gamma) ,  \quad (5)$$

where

$$Z(\beta,\gamma)=\int exp\{-\beta E-\gamma\Gamma\}dz=\iint dEd\Gamma\omega(E,\Gamma)exp\{-\beta E-\gamma\Gamma\}  \quad (6)$$

is the partition function, $\beta$ and $\gamma$ are Lagrange multipliers satisfying the equations for the averages

$$<E>=-\partial \ln Z/\partial \beta|_\gamma ;  \qquad <\Gamma>=-\partial \ln Z/\partial \gamma|_\beta .  \quad (7)$$

Introducing $\Gamma$ means effective account for more information than merely in linear terms of the canonical distribution $exp\{-\beta E\}/Z(\beta)$. The distribution (5) with the lifetime contains two different time scales: the first relates to the energy $E$, and the second – to the lifetime itself $\Gamma$, this latter one accounts for large-scale time correlations and large-time changes in $E$ by means of a thermodynamically conjugate to the lifetime $\gamma$. The similar operation can be derived from *NSO*.

The value $\gamma$ thermodynamically conjugated to the lifetime is related to the entropy fluxes and entropy production which characterize the peculiarities of the nonequilibrium processes in an open thermodynamic system. If $\gamma=0$ and $\beta=\beta_0=(k_B T_{eq})^{-1}$, where $k_B$ is the Boltzmann constant, $T_{eq}$ is the equilibrium temperature, then the expressions (5-6) yield the equilibrium Gibbs distribution. One can thus consider (5-6) as a generalization of the Gibbs statistics to cover the nonequilibrium situation. Such physical phenomena as the metastability, phase transitions, stationary nonequilibrium states are known to violate the equiprobability of the phase space points. The value $\gamma$ can be regarded as a measure of the deviation from the equiprobability hypothesis. In general one might choose the value $\Gamma$ as a subprocess of some other kind as chosen above. The canonical Gibbs distribution is derived from the microcanonical ensemble as zeroth approximation on the interaction between the system and the environment The lifetime $\Gamma$ allows effective account of this interaction (analogously to the methods of McLennan [36] and that of Zubarev *NSO* [19]). The value of $\gamma=0$ characterizes equilibrium isolated systems, presenting thus an idealisation. If the detailed balance is satisfied and in equilibrium $\gamma\neq 0$.

For the method exposed formally the relation of the $\Gamma$ value (slave process) to the master process $E$ is essential. However the lifetime concept bears also a more profound physical meaning, unifying the Newtonian approach to the absolute time and the ideas of the time generating matter. The lifetime parameter bears features adherent to both ordinary dynamical variables (energy, particle number) and coordinate variables (time). Mathematically introducing lifetime means acquiring additional information regarding an underlying stochastic process, beyond merely knowledge of its stationary distribution, exploring the (stationary) properties of its slave process. The irreversibility appears there as a consequence of the assumption that lifetime exists and is finite, and the moments of birth and death of a system are observable and physically essential.

Let us underline the principal features of the suggested approach.

1. We introduce a novel variable $\Gamma$ which can be used to derive additional information about a system in the stationary nonequilibrium state. We suppose that $\Gamma$ is a measurable quantity at macroscopic level, thus values like entropy which are related to the order parameter (principal macroscopic variable) can be defined. At the mesoscopic level the variable $\Gamma$ is introduced as a variable with operational characteristics of a random process slave with respect to the process describing the order parameter.

2. We suppose that thermodynamic forces $\gamma$ related to the novel variable can be defined. One can introduce the "equations of state" $\beta(<E>,<\Gamma>)$, $\gamma(<E>,<\Gamma>)$. Thus we introduce the mapping (at least approximate) of the external restrictions on the point in the plane $\beta, \gamma$.

3. We suppose that a "refined" structure factor $\omega(E,\Gamma)$ can be introduced which satisfies the condition $\int\omega(E,\Gamma=y)dy=\omega(E)$ (ordinary structure factor). This function (like $\omega(E)$) is the internal (inherent) property of a system. At the mesoscopic level we can ascribe to this function some inherent to the system (at given restrictions $(\beta_0,\gamma_0)$) random process. The structure factor has the meaning of the joint probability density for the values $E,\Gamma$ understood as the stationary distribution of this process. Provided the "reper" random process for the point $(\beta_0,\gamma_0)$, one can derive therefrom the shape of the structure function. If we model the dependence of the system potential of the order parameter by some potential well, the lifetime distribution within one busy period and probabilities $\omega(E,\Gamma)$ can be viewed as distributions of the transition times between the subset of the phase space (possibly of the fractal character) corresponding to the potential well, and the subset corresponding to the domain between the "zero" and the "hill" of the potential where from the system will roll down to the zero state. To determine the explicit form of $\Gamma$ (at $(\beta_0,\gamma_0)$) the algorithm of the asymptotic phase coarsening of complex system is used (Section 3).

4. It is supposed that at least for certain classes of influences the resulting distribution has the form (5), (6), that is the change of the principal random process belongs to some class of the invariance leading to this distribution which explains how one can pass from the process in the reper point $(\beta_0,\gamma_0)$ (for example, in equilibrium when $\gamma=0$ and $\beta=1/k_BT$) to a system in an arbitrary nonequilibrium stationary state. The thermodynamic forces should be chosen so that the distribution lead to new (measurable) values of $(<E>,<\Gamma>)$.

## 3. Distribution for indisturbed lifetime

The use of a specific lifetime distribution depends on the physical peculiarities of the system under investigation. For simple system with a single class of ergodic states (for example, definite volume cell with gas), the limiting exponential distribution is valid, which has been proved in works of Stratonovich [9], and on the level of pure mathematics – by Korolyuk and Turbin [37] et al. Such distribution is widely used in the theory of reliability and other sciences. By means of the exponential distribution the authors of [5-7] described the thermal and mass conduction, and chemical reactions. However complex systems with several classes of ergodic states and complicated structure of the state space (in general it can acquire fractal shape) are characterized by so-called homeokinetic plateau, whose lifetime distribution can be of various forms. In [18] it is suggested for the *NSO* method to use instead of exponential the gamma-distributions, as well as other functions accounting for physics of the systems.

Further detailed elaboration will require the concrete definition of the $\Gamma$ distribution and the interpretation of the Lagrange parameters $\beta$ and $\gamma$. The Lagrange parameter $\beta$ is supposed to be (likewise the equilibrium Gibbs statistics)

$$\beta=1/k_BT, \tag{8}$$

where $T$ is the average (over the body volume) local equilibrium temperature. Since at fluxes $\vec{q}\neq 0$ the temperature is not the same all over the bulk of the body, one can define $T$ in a system with volume $V$ as the volume average, i.e. $T=V^{-1}\int_V T(r,t)dr$ (the same definition was used in [35]). Of coarse, the thermodynamic description itself is supposed to be already coarse-grained. To get the explicit form of the $\Gamma$ distribution we shall use the general results of the mathematical theory of phase coarsening of the complex systems [37], which imply the following density of distribution of the lifetime for coarsened random process (see also [9]):

$$p(\Gamma<y)=\Gamma_0^{-1}exp\{-y/\Gamma_0\} \tag{9}$$

for one class of the stable states and the Erlang density

$$p(\Gamma<y)=\sum_{k=1}^{n} R_k\Gamma_{0k}^{-1}exp\{-y/\Gamma_{0k}\}; \qquad \sum_{k=1}^{n} R_k=1 \qquad (10)$$

in the case of several ($n$) classes of the ergodic stable states, $R_k$ – probability that the system is in the $k$–th class. The values $\Gamma_0$ and $\Gamma_{0i}$ are averaging of the residence times and the degeneracy probabilities over stationary ergodic distributions (in our case - Gibbs distributions). The physical reason for the realization of the distribution in the form (9-10) is the existence of the weak ergodicity in a system. Mixing the system states at big times will lead to the distributions (9-10). As we note in Section 2, the structure factor $\omega(E,\Gamma)$ has a meaning of the joint probability density of values $E,\Gamma$. For the distributions (9), (10) the functions $\omega(E,\Gamma)$ from (4), (6) take on the form:

$$\omega(E,\Gamma)=\omega(E)\Gamma_0^{-1}exp\{-\Gamma/\Gamma_0\}; \qquad \omega(E,\Gamma)=\omega(E)\sum_{k=1}^{n} R_k\Gamma_{0k}^{-1}exp\{-\Gamma/\Gamma_{0k}\} \qquad (11)$$

for the relations (9) and (10) respectively. Substituting into the partition function yield

$$Z(\beta,\gamma)=Z(\beta)(1+\gamma\Gamma_0)^{-1}; \qquad Z(\beta,\gamma)=Z(\beta)\sum_{k=1}^{n} R_k(1+\gamma\Gamma_{0k})^{-1} \qquad (12)$$

for (9) and (10), where $Z(\beta)=\int\omega(E)exp\{-\beta E\}dE$ is the Gibbs partition function. For $n=2$, $Z(\gamma)=Z(\beta,\gamma)/Z(\beta)=R_1/(1+\gamma\Gamma_{10})+R_2/(1+\gamma\Gamma_{20})$; $\Gamma_{k0}=1/(-s_k)$, $k=1,2$, where $s_{1,(2)}=-(\Lambda_{11}+\Lambda_{22})/2+(-)[(\Lambda_{11}+\Lambda_{22})^2/4-\Lambda_{11}\Lambda_{22}+\Lambda_{12}\Lambda_{21}]^{1/2}$; $R_1=(\Gamma_{20}-\Gamma_{10}\Gamma_{20}b)/(\Gamma_{20}-\Gamma_{10})$; $R_2=(\Gamma_{10}-\Gamma_{10}\Gamma_{20}b)/(\Gamma_{10}-\Gamma_{20})$; $b=\mu_1(\Lambda_{22}+\Lambda_{21})+\mu_2(\Lambda_{11}+\Lambda_{12})$; $\mu_1+\mu_2=1$, $\Lambda_{kj}$ are the transition intensities between the classes of states; $\mu_k$ is the probability of the initial state belonging to the $k$-th ergodicity class.

We have from (6)-(7) and (11)-(12) when $\Gamma_\gamma=-\partial lnZ(\beta,\gamma)/\partial \gamma|_\beta$; $\Gamma_0(V)=\Gamma_\gamma(V)|_{\gamma=0}$;

$$\Gamma_\gamma=\Gamma_0/(1+\gamma\Gamma_0), \qquad \gamma=1/\Gamma_\gamma-1/\Gamma_0, \qquad (13)$$

that is $\gamma$ is the difference between the inverse lifetimes of the open system $1/\Gamma_\gamma$ and the system without external influences $1/\Gamma_0$ which can degenerate only because of its internal fluctuations. The value $\gamma$ is thus responsible for describing the interaction with the environment and its existence is the consequence of the open character of a system. When defining $\gamma$ one should take into account all factors, which contribute to the interaction between the system and the environment. If one denotes in (13) $x=\gamma\Gamma_0$, then $x=x_1+x_2+...+x_n$, where the value $x_i$ is determined by the flux labelled by the index $i$. From expressions (3) and (5) - (6) it is visible, that the value $\gamma$ is connected to the entropy production $\sigma$. From comparison with the Extended Irreversible Thermodynamics [38] it is possible to show that $x=\gamma\Gamma_0\approx q_a t_{0a}/aR$, where $q_a$ are characteristic currents in the system, $a$ is density of values $a$ which is transferred by a currents $q_a$, $R$ is the size of the system, $t_{0a}$ is time of degeneration of the system, time for which the system will pass with homeokinetic plateau in a degeneration state. Time $t_{0a}$ is expressed through times of a relaxation of currents $\tau_q$ (for example, for neutron system in a nuclear reactor with a current of neutrons $\Phi$, $\tau_\Phi=1/v_n\Sigma_{tr}$, where $v_n$ is average speed of neutrons, $\Sigma_{tr}$ is transport section) and Onzager's factors $L_q$. For example, for the case of chemical reactions accompanied by diffusion $t_{0k}=t_{0k\ diff}=c_k R (\tau_{\rho diff}/k_B L_k)^{1/2}$, where $c_k$ is concentration (density) of particles – an analogue of $a$, $\tau_{\rho diff}$ is characteristic correlation time of the diffusive fluxes $\vec{J}_k$; $J_k$ are projections of $\vec{J}_k$ (flux entering the system of the size R) on the external surface normal; $L_k = D_k T/(\partial \mu_k/\partial c_k)_{T,P}$; $\mu_k$ is chemical potential ($=\mu_{0k}$ without perturbation); $L_k$ is Onsager coefficients from $\vec{J}_i = -L_i \vec{\nabla}(\mu_i/T)_{T,P}$, $D_k$ the diffusion coefficient of the $k$-th component. If in the system the sources of value $a$ are present, then $x_a = y_a t_{0a}$, $y_a = (q_a - R\sigma_a)/Ra$, where $\sigma_a$ is density of a source of value $a$.

Let's note that the value similar to $\gamma$ it is defined in [39-41] for fractal objects. It is equal to zero for the closed system, and for open system it is equal to $\Sigma \lambda_i - \lambda_{KS}$, where $\lambda_i$ are Lyapunov's parameters, and $\lambda_{KS}$ is Kolmogorov-Sinai entropy.

## 4. Superstatistics from distribution of the kind (5)

In the distribution (4)-(5) containing lifetime, as thermodynamic parameter, joint probability for values $E$ and $\Gamma$ is equal

$$p(E,\Gamma) = \frac{e^{-\beta E - \gamma \Gamma} \omega(E,\Gamma)}{Z(\beta,\gamma)}; \quad Z(\beta,\gamma) = \int e^{-\beta E - \gamma \Gamma} dz = \iint dE \, d\Gamma \, \omega(E,\Gamma) e^{-\beta E - \gamma \Gamma}. \quad (14)$$

Having integrated (14) on $\Gamma$, we obtain distribution of a kind

$$p(E) = \int P(E,\Gamma) d\Gamma = \frac{e^{-\beta E}}{Z(\beta,\gamma)} \int_0^\infty e^{-\gamma \Gamma} \omega(E,\Gamma) d\Gamma. \quad (15)$$

According to the third assumption from Section 2, the structural factor $\omega(E,\Gamma)$ is meaningful to joint probability for $E$ and $\Gamma$ (11), treated as stationary distribution of this process. We shall write down

$$\omega(E,\Gamma) = \omega(E)\omega_1(E,\Gamma) = \omega(E) \sum_{k=1}^n R_k f_k(\Gamma, E). \quad (16)$$

In last equality (16) it is supposed, that there exists $n$ classes of ergodic states in a system; $R_k$ is the probability of that the system will be in $k$-th a class of ergodic states, $f_k(\Gamma, E)$ is density of distribution of lifetime $\Gamma$ in this class of ergodic states (generally $f_k$ depends from $E$). As physical example of such situation (characteristic for metals, glasses) one can mention the potential of many complex systems of a kind, for example, *Fig. 2*. Such situation is considered in [42]. Minima of potential correspond to the metastable phases, disproportionate structures, etc. Essential object of research of statistical physics recently became complex nonergodic systems: the spin and structural glasses, disorder geteropolimery, the granular media, transport currents, etc. [43]. The basic feature of such systems will be, that their phase space is divided into isolated areas, each of which corresponds to a metastable thermodynamic state, and the number of these areas exponential exceeds full number (quasi)particles [44]. The quasithermodynamical theory of structural transformations of alloy *Pd-Ta-H*, based on this model, is constructed in [45]. Expression (16) to the description of such systems is applicable.

We shall assume an obvious kind of distribution $f_k$ in (16), having chosen it as gamma-distribution

$$f_k(x) = \frac{1}{\Gamma(\alpha_k)} \frac{1}{b_k^{\alpha_k}} x^{\alpha_k - 1} e^{-x/b_k}, \quad x > 0, \quad f_k(x) = 0; \quad x < 0; \quad \int_0^\infty e^{-\gamma_k x} f_k(x) dx = (1 + \gamma_k b_k)^{-\alpha_k}. \quad (17)$$

($\Gamma(\alpha)$ is gamma-function). Substituting (17) in (15)-(16), we receive, that

$$p(E) = \frac{e^{-\beta E}}{Z(\beta, \gamma)} \omega(E) \int_0^\infty \sum_{k=1}^n R_k f_k(\Gamma|E) e^{-\gamma \Gamma} d\Gamma = \frac{e^{-\beta E}}{Z(\beta, \gamma)} \omega(E) \sum_{k=1}^n R_k (1 + \gamma_k b_k)^{-\alpha_k}. \quad (18)$$

From expression (7):

$<\Gamma_\gamma> = \alpha b/(1+\gamma b); \Gamma_0 = <\Gamma_\gamma>_{|\gamma=0} = \alpha b; 1+\gamma b = \Gamma_0/<\Gamma_\gamma>; (1+\gamma_k b_k)^{-\alpha_k} = exp\{-\alpha_k ln(\Gamma_0/<\Gamma_\gamma>)\}.$ (19)

At a conclusion (19) we assumed weak dependence of $<\Gamma_\gamma>$ from $E$ though below this dependence is essentially considered. The value $\Gamma_{0k}$ is equal to average lifetime in $k$-th a class of metastable states without disturbance from (9)-(13). We understand absence of stationary forces and the determined fluxes $f = -<I_v>$ (A.3) as indisturbed lifetime $\Gamma_{0k}$; at system (open metastable area) there are only stochastic fluxes $\rho$. For average lifetime $\Gamma_{0k}$ of the system in dynamical equilibrium (or in stationary) state, where at system with grand canonical sum $Q_k$ are only stochastic fluxes $\rho$, in work [32] (see also Appendix A) by means of stochastic models of storage it is obtained

$$<\Gamma_{ok}> = \frac{1}{\lambda_k}(Q_k - 1), \quad (20)$$

where $Q_k = exp\{\beta_k P_k V_k\}$ is the grand statistical sum of the grand canonical ensemble of the part of system in $k$-th a metastable state, in the $k$-th potential well, $\beta_k = (1/k_B T)_k$ is reverse temperature in $k$-th a metastable state, $P_k$, $V_k$, $T_k$ are pressure, volume, temperature in $k$-th a metastable state, $V$ is the full system volume, $\lambda_k$ is intensity of energy flow in the system (subsystem), equal in dynamical equilibrium of an output intensity [32]. For realistic metastable systems the detailed balance principle is satisfied and $\lambda \neq 0$. For example, $1\ cm^2$ of the surface of an equilibrium water drop evaporates $10^{21}$ molecules per second; the same amount per time unit undergoes condensation. .

Using (20) in (19), we receive:

$exp\{-\alpha_k ln(\Gamma_0/<\Gamma_\gamma>)\} = exp\{-\alpha_k[ln((exp\{(\beta PV)_{0k}\}-1)/(exp\{(\beta PV)_{\gamma k}\}-1)) + ln(\lambda_{\gamma k}/\lambda_{0k})]\}.$

For not so small systems, the number of particles in which is more, than $3$-$5$, $exp\{\beta_k P_k V_k\} >> 1$. Then

$d = (exp\{(\beta PV)_{0k}\}-1)/(exp\{(\beta PV)_{\gamma k}\}-1) \approx exp\{(\beta PV)_{0k} - (\beta PV)_{\gamma k}\} = exp\{[\beta_{0k} P_{0k} v_{0k} - \beta_{\gamma k} P_{\gamma k} v_{\gamma k}]E/u\},$
where $u = E/V$ is specific energy, $v_k = V_k/V$, $v_{0k} = V_{0k}/V$; $v_{\gamma k} = V_{\gamma k}/V$.

If to enter average values $<v> = \sum_{k=1}^n R_k v_k$, $<P> = \sum_{k=1}^n R_k P_k$, $<\beta> = \beta^0 = \sum_{k=1}^n R_k \beta_k$, we shall receive:

$d=\exp\{y_k E\Delta/u\}$, $y_k=[\beta_{0k}P_{0k}v_{0k}-\beta_{\gamma k}P_{\gamma k}v_{\gamma k}]/\Delta$, $\Delta=<\beta Pv>_0-<\beta Pv>_\gamma$, $<y_k>=1$. Then

$$\sum_{k=1}^{n} R_k(1+\gamma_k b_k)^{-\alpha_k} \approx \sum_{k=1}^{n} R_k \exp\{-\alpha_k y_k E\Delta/u - \alpha_k \ln(\lambda_{\gamma k}/\lambda_{0k})\}.$$

The second composed does not depend from $E$ and joins in normalizing multiplier $Z$ in expression

$$p(E) = \frac{e^{-\beta E}}{Z(\beta,\gamma)} \omega(E) \sum_{k=1}^{n} R_k e^{-r_k E}, \qquad (21)$$

received from (18)-(20), here $r_k=\alpha_k y_k \Delta/u = \alpha_k[\beta_{0k}P_{0k}v_{0k}-\beta_{\gamma k}P_{\gamma k}v_{\gamma k}]/u$; $<r_k>=r_0=\alpha_0\Delta/u$; $\alpha_0 = \sum_{k=1}^{n} R_k \alpha_k$.

The unknown value remains $\alpha$. From distribution (17) we shall find, that $\alpha=<\Gamma>^2_0/(<\Gamma^2>_0-<\Gamma>^2_0)$. We find the value $<\Gamma^2>_0$ from stochastical model of storage [32] (Appendix A)

$E(\exp\{-s\Gamma_x\})=\exp\{-x\eta(s)\}$; $\eta(s)=s+\varphi(\eta(s))$; $<\Gamma^2>_0-<\Gamma>^2_0=\rho\sigma^2/\lambda(1-\rho)^3$; $\sigma^2=\int_0^\infty x^2\lambda b(x)dx$;

$<E>_{st}=\sigma^2 e/2(1-\rho)t_0$; $\alpha=(1-1/Q)e/2<E>_{st}\lambda t_0 \approx e/2<E>_{st}\lambda t_0$; $<\alpha>=\sum_{k=1}^{n} R_k/2<E_k>_{st}\lambda t_0 \approx$

$e/2<E>_0\lambda_0 t_0$; $\lambda_0 = \sum_{k=1}^{n} R_k \lambda_k$; $<E>_0 = \sum_{k=1}^{n} R_k <E_k>$,

where $t_0$ is average time of an output from system of one particle, $e$ is average energy of one particle of system, $<E_k>$ - is average energy of the system in $k$-th a metastable state. Thus $r_0=\Delta e/u2<E>_0\lambda_0 t_0$. For exponential disturibution $f_k(x)$ in (17) $\alpha_k=1$, $2<E>_k\lambda_k t_0/e=1$. In many cases instead of gamma-distribution (17) it is possible to use exponential distribution with $\alpha_k=1$. Then $r_0=\Delta/u$. At $\gamma=0$, $r_0=0$, expression for $\gamma$ equal

$\gamma=(\exp\{-(\beta PV)_\gamma\}-\exp\{-(\beta PV)_0\})/2<E>_k(t_0/e)$.

By $\alpha_k=1$, $\gamma=\lambda[\exp\{-(\beta PV)_\gamma\}-\exp\{-(\beta PV)_0\}]$; $r_k E=\ln[1+(\gamma_k/\lambda_k)\exp(\beta PV)_{0k}]$. For simplicity we shall consider a case of $\alpha=1$ though it is simple to consider and more the general case $\alpha\neq 1$.

By $\alpha=1$

$$p(E) = \frac{e^{-\beta E}}{Z(\beta,\gamma)} \omega(E) \sum_{k=1}^{n} R_k \frac{<\Gamma_\gamma>}{<\Gamma_0>} = \frac{e^{-\beta E}}{Z(\beta,\gamma)} \omega(E) \sum_{k=1}^{n} R_k [1 + \frac{\gamma_k}{\lambda_k} \exp\{(\beta PV)_{0k}\}]^{-1}.$$

This distribution tends to Gibbs one by $\gamma\to 0$. By big values $\gamma$, $p(E)\to 0$. By $\gamma\approx\lambda$,

$p(E) \sim \frac{e^{-\beta E}}{Z(\beta,\gamma)} \omega(E) \sum_{k=1}^{n} R_k \exp\{-(\beta Pv)_{0k} E/u\}$.

As an example of value $\gamma/\lambda$ it is possible to consider this value for a random phase of the synchronizable generator [12], when

$<\Gamma_\gamma>=exp(\pi D_0)|I_{iD0}(D)|^2/(1+exp(2\pi D_0))|I_0(D)|^2\lambda$, $\Gamma_0=1/2\lambda$,

where $D_0$ connected with initial difference of frequencies of synchronizable generators and with disturbance $\gamma$, parameter $D$ characterizes size of the attitude a signal-noise in a strip of synchronization and it is connected with width of a strip of synchronization (deduction) $\Delta$, $\lambda=N^{+/-}|_{D0=0}=\Delta/4\pi D^2|I_0(D)|^2$ is intensity of phases jumps at $D_0=0$, $I_{iD0}(D)$ are tabulated Bessel's functions of imaginary argument and imaginary index. Then

$$\gamma/\lambda=(1/\lambda)(1/<\Gamma_\gamma>-1/<\Gamma_0>)=[1+exp(2\pi D_0)]|I_0(D)|^2/exp(\pi D_0)|I_{iD0}(D)|^2-2=\pi D_0[exp(\pi D_0)+exp(-\pi D_0)]/sh\pi D_0[1+2D_0^2 I^2_0(D)\sum_{n=1}^{\infty}(-1)^n I_n(D)/(n^2+D_0^2)]-2.$$

This value tends to $0$ at $\gamma\to 0$ and tends to $\infty$ as $D_0$ at $D_0\to\infty$ (though $D_0$ have a finite values).

At such an approach which can be considered as a discretized version of superstatistics [1, 2], one accounts for not only the inverse temperature fluctuations (that of $\beta$), but the fluctuations of values $P, v, \alpha$ as well.

Denote in (21) $r_k=y_k\Delta/u=y_k r_0$; $r_0=\Delta/u$ and rewrite the exponent under sum in rhs of (21) as $y_k r_0 E$. It will be shown below that $r_0$ is related to the controlling parameter of a problem (for example, with the feedback coefficient in the Van der Pol generator, the birth parameter in the Maltus-Ferhuelst process etc). Since $R_k$ is the probability of that the system will be in $k$-th a class of ergodic states, then $R_k$ is the probability of that the quantity $y=(r/r_0)$ will take on value $y_k=(r_k/r_0)$; $R_k=R(y=y_k)$. Thus the expression under sum in rhs of (21) is the characteristic function $\varphi(t)=\sum_{i=1}^{n}e^{itx_i}R_i$ of the distribution $R_k$ with argument $it=-r_0E$. If $R_k=\delta_{k0}$, where for a cell with zero index $r_k=r_0$, then (21) acquires the form $p(E)\sim exp\{-\beta E-r_0 E\}=exp\{-\beta^0 E\}$, where the open character of the system (existence of fluxes) is taken into account, therefore the inverse temperature is set to be not $\beta=1/kT$, but $\beta^0$. Thus the relation between $\beta$ and $\beta^0$ is $\beta_0=\beta+r_0$; $\beta=\beta^0-r_0$. If $\beta^0$ weak depended from $\gamma$, then $r_0\approx\beta^0 p_0/u$; $p_0=<Pv>_0-<Pv>_\gamma$; $\beta=\beta^0(1-p_0/u)$; $\beta^0=\beta/(1-p_0/u)$.

We shall write down (21) in the form

$$p(E)=f_B(E)\varphi(E)/Z=p_B p_A(Z_B Z_A/Z); \quad f_B(E)=exp\{-\beta E\}\omega(E); \quad \varphi(t)=\sum_{i=1}^{n}e^{itx_i}R_i ; \quad (22)$$

$it=-r_0 E; \quad Z=\int f_B\varphi(E)dE; \quad Z_A=\int\varphi(E)dE; \quad Z_B=\int f_B dE; \quad p_B=f_B/Z_B; \quad p_A=\varphi(E)/Z_A$,

and consider different resources of task of function $\varphi(E)$ in (22). Let us choose as $R_k$ in (22) the probabilities of the negative binominal distribution (a discrete analogue of the continuous gamma distribution) with characteristic function $\varphi(t)=\sum_{i=1}^{n}e^{itx_i}R_i=[p/(1-(1-p)exp\{it\})]^s$, $s$ and $p$ being parameters. Setting $s=1/(q-1)$, $it=-r_0 E$, $p=1/[1+(q-1)]=1/q$, we conclude that (22) in this case is

$$p(E)\sim exp\{-\beta E\}[1+(q-1)(1-exp\{-r_0 E\})]^{-1/(q-1)}. \quad (23)$$

The same result can be written for the Polia distribution characterized by the characteristic function $\varphi(t)=[1+\alpha\lambda(1-exp\{it\})]^{-1/\alpha}$ at $\alpha=q-1$, $\lambda=1$. In this approach the behaviour of $p(E)$ is asymptotically not power-like, as in Tsallis [3, 4] distributions, but rather exponential. The argument $r_0 E$ in the Tsallis distribution ([3, 4], see also the part II) is replaced by $(1-exp\{-r_0 E\})$.

To find an asymptotics at big values of the argument by $exp\{-r_0E\}<<1$, lets develop into series the second factor in the rhs of (23):

$$p(E)\sim exp\{-\beta E\}[1+exp\{-r_0E\}/q+exp\{-2r_0E\}/2q+(2q-1)exp\{-3r_0E\}/6q^2+...].$$

As $R_k$ one can choose various specific forms of distributions. The negative binominal distribution can be considered as a distribution of a random value with Poisson distribution, whose parameter $\mu$ is itself a random value distributed via the gamma-law with the parameters $\lambda=p/(1-p)$, $\alpha=r$ [46]. Poisson distribution at independent trials describes the trapping by the $k$-th potential well. Real trapping by the $k$-th well is affected by many factors which is reflected in the random character of the parameter $\mu$. One more way of arriving at the negative binominal distribution [46] is considering the fact that it is the particular case of the generalized Poisson distribution. It describes the sum $S_N=X_1+...+X_N$ of the random number $N$ of independent random $X_i$, described by the characteristic function $f(t)$, the random number $N$ being Poisson distributed with the parameter $\lambda$. If for $f(t)$ one chooses the characteristic function of the logarithmic distribution [46] of the kind $f(t)=ln[1-(1-p)exp\{it\}]/lnp$, from the general expression for the characteristic function of the generalized Poisson distribution $\varphi(t)=exp\{-\lambda+\lambda f(t)\}$ taking $s=-\lambda/lnp$, we get the characteristic function $[p/(1-(1-p)exp\{it\})]^s$ of the negative binominal distribution. If for the generalized Poisson distribution we choose as $f(t)$ the characteristic function of the negative binominal distribution with the above defined parameters, we get for the distribution function (22)-(23):

$$p(E)\sim exp\{-\beta E-\lambda+\lambda f(exp\{-r_0E\})\}\sim exp\{-\beta E-\lambda+\lambda[1+(q-1)(1-exp\{-r_0E\})]^{-1/(q-1)}\}. \quad (24)$$

We define the $\lambda$ parameter, likewise it was done above, $\lambda=-slnp=ln[1+(q-1)]/(q-1)$. Thereby $<r_k>\neq r_0$; $<r_k>=\lambda r_0$. One can obtain the equality $<r_k>=r_0$ choosing the parameter $p$ of the characteristic function of the negative binominal distribution (24) as $p=(1+(q-1)/\lambda)^{-1}=(1+(q-1)^2/lnq)^{-1}$ (the last equality corresponds to our choice of the $\lambda$ parameter). Thus instead of (24) we have:

$$p(E)\sim exp\{-\beta E-\lambda+\lambda[1+(q-1)^2(1-exp\{-r_0E\})/lnq]^{-1/(q-1)}\}.$$

Other values of $\lambda$ can be equally used. In this framework the system falling into the $k$-th valley is described by a random sum $X_i$ of the clusters of the valley [42, 45]. The number of valleys in a single cluster is distributed over the negative binominal distribution with $s=1/(q-1)$, $p=1/[1+(q-1)]$. At $q\rightarrow 1$ the distribution (24) passes to $p(E)\sim exp\{-\beta E-1+exp\{-(1-exp\{-r_0E\})\}\}$.

In the absence of fluxes and $r_0\rightarrow 0$ we get the Gibbs distribution $p(E)\sim exp\{-\beta E\}$. The same distribution can be obtained from (23) at $r_0\rightarrow 0$. If we expand (24) over $\lambda$, then $p(E)\sim exp\{-\beta E\}[1+\lambda([1+(q-1)(1-exp\{-r_0E\})]^{-1/(q-1)}-1)+\lambda^2[1+(q-1)(1-exp\{-r_0E\})]^{-1/(q-1)}-1)^2/2+...]$, which presents Gibbs distribution and its amendments.

One more possibility of obtaining more detalized distributions lies in the use of compound characteristic functions, when one of the functions serves as argument for another: $\varphi(f(t))$. The probabilistic interpretation of such relation is that they correspond to the $S_N=X_1+...+X_N$ of random number $N$ of independent random values $X_i$. Physically they describe hierarchical systems; such hierarchical fractal structure of potential profiles plays important part in realistic complex systems [42, 45]. One can use several levels of hierarchy: $\varphi(f_1(f_2(...)))$. The use $\varphi(f(t))$ of the negative binominal distribution and expressions like (23) $\varphi$ and $f$ leads to the $p(E)$ of the kind

$$p(E))\sim exp\{-\beta E\}\{1+(q-1)(1-[1+(q_1-1)(1-exp\{-r_0E\})]^{-1/(q_1-1)})\}^{-1/(q-1)}.$$

The use of $\varphi(f_1(f_2(t)))$ and the negative binominal distribution for $\varphi, f_1, f_2$ yields

$$p(E))\sim exp\{-\beta E\}\{1+(q-1)(1-[1+(q_1-1)(1-[1+(q_2-1)(1-exp\{-r_0E\})]^{-1/(q_2-1)})]^{-1/(q_1-1)})\}^{-1/(q-1)}.$$

## 5. Examples of application of expressions (5), (14) with a lifetime

As concerning a simple example of the stationary nonequilibrium state described by distributions (5) - (7), (14) - (16), we shall consider a flow of rarefied gas in a tube of radius $R$ and lengths $L$, $L>>R$. In this system there is one class of ergodic states. If in expression (15) to choose structure factor $\omega(E,\Gamma)$ as (9), (11) we obtain:

$$p(E)=exp\{-\beta E\}\omega(E)/Z(1+\gamma\Gamma_0). \tag{25}$$

From (13) we find, that $1/(1+\gamma\Gamma_0)=<\Gamma>/\Gamma_0$, where $<\Gamma>$ is average a lifetime of the system. Now for the determination of $<\Gamma>$ we apply the stochastic model of storage with constant exit rate $r=a_f(1-\chi_{Z(t)})$ which has been used in work [32] (Appendix A).

The random rate of change of number of elements $Z$ in this model is described by the stochastic equation $dZ/dt=dA_f/dt-a_f(1-\chi_{Z(t)})$, where $\chi_{Z(t)}=0$, $Z(t)>0$; $\chi_{Z(t)}=1$, $Z(t)=0$; $dA_f/dt$ is rate of random input of elements in the system, $P_{Z0}(0,t)=E\{\chi_{Z(t)}\}=P\{Z(t)=0|Z(0)=Z_0\}$; $P_0=lim_{|t\to\infty}P_{Z0}(0,t)$ is the stationary probability of degeneration. As in (A.3), Appendix A, there act the stationary external forces $v$ supporting difference of pressure on the ends of the tube, and in their role enter the walls of a tube reflecting particles of rarefied gas, collisions between which we neglect. We get from (A.3) that $<\Gamma>=<x_f>/a_f(1-\rho_f/a_f)$, where $<x_f>=\int xb_f(x)dx$ is average size of the input "batches", $b_f(x)$ is density of distribution of size the input "batches"; $\rho_f=\rho-f$, $\rho=\lambda<x>$, $\lambda$ is intensity of an inputing flux, $-f = <I_v>$ is the average flux connected to enclosed force $v$. Then

$$1/(1+\gamma\Gamma_0)=(<x_f>/a_f\Gamma_0)/(1-\rho_f/a_f). \tag{26}$$

The multiplier $<x_f>/a_f\Gamma_0$ is included in normalizing multiplier $Z$. It is possible to show that stationary value of a flux $q$ in this system is equal to $\rho_f$, as $q=a_f(1-P_0)=\rho_f$, $P_0=1-\rho_f/a_f$. For rarefied gas the flux in a tube is $q=2R(m/2k_BT)^{1/2}(P_1-P_2)/3\pi L$ [49], where $(P_1-P_2)$ is the difference of pressure on the ends of a tube, $m$ is weight of atoms, $T$ is temperature. This expression for a flux is obtained from Maxwell-Boltzmann distribution, and can be considered as taken in zero order. Average value of internal energy, thermodynamic internal energy $U$ for ideal gas it is $U=C_VT$, where $C_V$ is a thermal capacity of system at constant volume [49]. We assume a constancy of temperature in a tube equal to temperature of walls. The pressure is $P=nk_BT/m$, where $n$ is density of gas; then $q=\rho_f=2R(1/2m)^{1/2}(n_1-n_2)(k_BU/C_V)^{1/2}/3\pi L$. If in this expression and in (26) we replace average value of a flux $q$ with microscopic value of a flux having put in $q$, $U=E$, where $E$ is microscopic random internal energy (this can be done since real flux and energy in system are random variables), from (25) - (26) the distribution for energy becomes

$$p(E)=exp\{-\beta E\}\omega(E)/Z[1-q(E)/a_f]=$$
$$exp\{-\beta E\}\omega(E)/Z[1-2R(1/2m)^{1/2}(n_1-n_2)(k_BE/C_V)^{1/2}/3\pi La_f]. \tag{27}$$

At $\rho_f \geq a_f$ in system there are no stationary states and stationary distributions. Value $a_f=const$ can be defined as the maximal value of a flux, at which there exist stationary states in laminar mode. If due to fluctuations of energy $E=\sum_{k=1}^{N} m(v_k-u)^2/2$ ($v_k$ is velocity of $k$-th particle, $u$ is average value of velocity) or of density $n_1$, $n_2$, value of a random flux $q$ will exceed the value $a_f$, there arise arises a turbulent mode in a system, new structures for which a new stationary distribution is formed. Value $q=a_f$ is a point of nonequilibrium phase transition from a laminar mode to turbulent.

The obtained distribution (27) can be written and in the other form if if we multiply and divide by value $q(E)=2R(1/2m)^{1/2}(n_1-n_2)(k_BE/C_V)^{1/2}/3\pi L$ on density in the middle of a tube $n(L/2)=<n(E)>=\int n(E)p(E)dE$ and assume that $n(L/2)\approx n(E)=p(E)n_0$, $n_0=\int n(E)dE$. Performing this for $n(L/2)$ in numerator we get a quadratic equation for $p(E)$ with the solution $p(E)=[1+(1-(qn_0/a_fn(L/2))4exp\{-\beta E\}\omega(E)/Z)^{1/2}]/2(qn_0/a_fn(L/2))$. The minus sign is chosen because at $q\to 0$, $p(E)\to exp\{-\beta E\}\omega(E)/Z$, as in (27). At $q/a_f\geq 1$ expression under radical can take negative values, and as in (27), the stationary mode does not exist. However, in this case the condition of breaking the stationarity is expressed not so precisely, as in (27). Decomposing square root into series we get that $p(E)=(exp\{-\beta E\}\omega(E)/Z)[1+(exp\{-\beta E\}\omega(E)/Z)(qn_0/a_fn(L/2))+2(exp\{-\beta E\}\omega(E)/Z)^2(qn_0/a_fn(L/2))^2+...]$. Putting $n(L/2)\approx n(E)=p(E)n_0$ in a denominator of expression for $q$ in (27) yields $p(E)=exp\{-\beta E\}\omega(E)/Z+q(E)n(L/2)/a_fn_0$. At $n(L/2)/n_0\approx exp\{-\beta E\}\omega(E)/Z$ the expression for $p(E)$ is $p(E)\approx(exp\{-\beta E\}\omega(E)/Z)[1+q(E)/a_f]$; this expression can be considered as the first member of decomposition (27) at small values $q/a_f$, and does not take into account phase transition, when $q/a_f\to 1$.

As one more example we shall consider system of neutrons in a nuclear reactor (*NR*) and we shall obtain distribution of energy of neutrons in a nuclear reactor in view of finiteness of their lifetime. All characteristics of a nuclear reactor are averaged on distribution of energy of neutrons in a reactor (or on a flux of neutrons depending on energy), therefore obtaining these distributions is important. Among known approximations are Maxwell distribution for thermal neutrons and Fermi's distribution for fast neutrons, and intermediate neutrons are described by various approximations. In the present section a uniform distribution for all energy range is obtained. In [50] it is marked that the equilibrium condition of neutrons with nucleus of moderator can not be reached because of finiteness of lifetime of neutrons. For neutrons with energy above energy of thermal group the spectrum differs from Maxwell one.

As classes of ergodic states are considered prompt neutrons (a zero class) and $n$ groups of the delayed neutrons which are emitted by corresponding precursors, fission fragment isotopes. From (16) - (17)

$$p(E)=exp\{-\beta E\}Z^{-1}\omega(E)\sum_{k=0}^{n}R_k/(1+\gamma_k b_k)^{-\alpha_k}. \qquad (28)$$

If $f_k$ in (16) - (17) is exponential distribution (9-12) then $\alpha_k=1$, $b_k=\Gamma_{0k}$. For exponential distribution and one class of ergodic states, when $n=0$, $R_k=\delta_{ko}$, $\Gamma_{0k}=\Gamma_0$, $\gamma_k=\gamma$, $(1+\gamma_k b_k)^{-\alpha_k}=(1+\gamma\Gamma_0)^{-1}=<\Gamma>/\Gamma_0$ (13), where $<\Gamma>$ is average lifetime of system, we have a situation when we take into consideration only prompt neutrons. In [18] it is shown that for neutrons in a nuclear reactor $<\Gamma>=-1/\omega$, where $-\omega=1/T$, $T$ is the period of a reactor; $<\Gamma>=l_{ef}/(1-k)$, [50-52], where $l_{ef}=1/v\Sigma_a\sim 10^{-3}$ s is average effective lifetime of a neutron in reactor ($v$ is the average velocity of neutrons, $\Sigma_a$ is the macroscopic absorption cross section of neutrons in reactor), $k$ is the effective reproduction factor of neutrons.

Assuming $\Gamma_0=<\Gamma>_{k=0}$, yields $\Gamma_0=l_{ef}$, $<\Gamma>/\Gamma_0=1/(1-k)$. Then

$$p(E)=exp\{-\beta E\}Z^{-1}\omega(E)/(1-k). \qquad (29)$$

At $k=1$ there are no stationary distributions and stationary states. This situation can be modeled by the stochastic theory of storage [48] in which the same conditions of the stationarity breaking take place . In case of $n$ groups of the delayed neutrons values $<\Gamma_i>=-1/\omega_i$ are looked as the solution of the inhour relationship [50], [51], [52]. For one group of the delayed neutrons it has a form

$$k\rho(\omega+\lambda)=\omega l_{ef}(\omega+\lambda)+\beta\omega k, \qquad (30)$$

where $\rho=(k-1)/k$ is the reactivity of a reactor, $\lambda=0,077\ s^{-1}$ is a decay constant of precursors of the delayed neutrons, $\beta=0,0065$ is the yield of delayed neutrons. Solving the equation (30), assuming $R_0=(1-\beta)$, $R_1=\beta$ are probabilities of the neutrons belonging to classes of prompt or delayed neutrons, accordingly, $\Gamma_{00}=l_{ef}$, $\Gamma_{01}=1/\lambda$ are undisturbed average lifetimes of prompt and delayed neutrons, we get from (28) that for one group of delayed neutrons

$$p(E)=exp\{-\beta E\}Z^{-1}\omega(E)[2(1-\beta)/(\lambda l_{ef}+\beta k+1-k)(1-C)+2\beta\lambda l_{ef}/(\lambda l_{ef}+\beta k+1-k)(1+C)]; \quad (31)$$
$$C=[1+4k\lambda l_{ef}(k-1)/(\lambda l_{ef}+\beta k+1-k)^2]^{1/2}.$$

In (29), (31) it is accepted $\omega(E)\sim E^{1/2}$. We shall return to expression (29) and we shall consider dependence $k(E)$. We take into account that fissions are caused by neutrons of arbitrary energy $E`$ and that neutrons of fission also possess a spectrum. Number $A(E)$ of neutrons of fission with energy $E$, generated in $1\ sm^3$ for $1\ s$, is described of expression [51]

$$A(E)=\int_0^\infty \Sigma_f(E`)\varphi(E`)v(E`,E)dE`\approx \overline{\Sigma}_f\overline{\varphi}\ n_f(E), \qquad (32)$$

where $\Sigma_f$ is the macroscopic fission cross section, $\varphi(E)=n(E)v(E)$ is density of a flux of neutrons; $v(E)\approx 1,38*10^{-4}E^{1/2}$ is the velocity of neutrons; assuming random character $E$, we shall write down function of distribution $n(E)=p(E)n_0$; $n_0=\int_0^\infty n(E)dE$. Function $v(E`,E)$ shows how many fission neutrons with energy $E$ arise in one act of fission caused by a neutron with energy $E`$, and function [51]

$$n_f(E)=\int_0^{0,1ev} v(E`,E)dE`=0,4527exp\{-E/0,965\}sh(2,29E)^{1/2}\approx 0,4527E^{1/2}exp\{-E/1,29\} \qquad (33)$$

in (32) means a spectrum of fission neutrons at fission into thermal neutrons. Values $\overline{\Sigma}_f, \overline{\varphi}$ in (32) designate average on $E$ (on a spectrum) values $\Sigma_f(E)$ and $\varphi(E)$. Then the reproduction factor for neutrons with energy $E$ is given by expression

$$k(E)=A(E)/\Sigma_{a\ full}^{tot}(E)\varphi(E), \qquad (34)$$

where $\Sigma_{a\ full}^{tot}\ (=\Sigma_a+\Sigma_f)$ is the total macroscopic absorption cross section (for nucleus of all types), including fission, $A(E)$ is given in (32) - (33).

Thus, substitution (32) - (34) in (29) in view of that $\varphi(E)=p(E)n_0v(E)$, leads to the equation for $p(E)$ which generally looks like the integral equation. However, the approach made in the second part of a expression (32), reduces this integral equation to the simple algebraic equation with the solution

$$p(E)=exp\{-\beta E\}Z^{-1}\omega(E)+\overline{\Sigma}_f\overline{\varphi}\ n_f(E)/\Sigma_{a\ full}^{tot}(E)n_0v(E)=f_B(Z_\beta/Z)+f_1(1-Z_\beta/Z), \qquad (35)$$

where $n_f(E)$ it is given in (33), $f_B=exp\{-\beta E\}Z_\beta^{-1}\omega(E)$ is a spectrum of thermal neutrons, $\beta \approx 1/k_B E_t \approx 30$ eV, $E_t$ is average energy of thermal neutrons, $k_B$ is Boltzmann constant, and $f_1 = C\bar{\Sigma}_f \bar{\varphi}\, n_f(E)/\Sigma_{afull}^{tot}(E)n_0 v(E)$ is the additive to Boltzmann distribution of the thermal neutrons $f_B$, considering features of behaviour of fast and intermediate neutrons; $Z_\beta=\int exp\{-\beta E\}\omega(E)dE$; $Z=\int exp\{-\beta E\}(1-k)^{-1}\omega(E)dE$; $\int f_B dE = \int f_1 dE = 1$. For dependence $\Sigma(E)$ far from resonances the law $\Sigma(E) \sim const/E^{1/2}$ is fair, and generally Breit-Wigner formula is fair [51]:
$$\Sigma(E)=constE^{-1/2}1/[(E-E_r)^2+\Delta^2/4],$$
where $\Delta$ is the natural energy width of the resonance, $E_r$ is the resonance peak. The solution of the integral equation leads to the same functional dependence (35). We shall note that expression (35) resembles a method in which distribution is represented in the form of the sum Maxwell and Fermi (with some correction multiplier) distributions [52], but the analytical form of second term is essentially different. However, the behaviour is similar. The dependences obtained from expression (35), are presented on fig. 3, where $p(E)=E^{1/2}exp\{-25E\}+0,02E^{1/2}exp\{-E/0,965\}sh(2,29E)^{1/2}$; $m(E)=E^{1/2}exp\{-25E\}$. Comparison of expression (35) with Maxwell distribution $m(E)$ is shown. Such behaviour $p(E)$ as on fig.3, corresponds to experimentally observable results ([52], fig. 6.7, 8.19, 8.20). Shown on fig. 8.20b [52] breaks of a curve describe compose resonanses in $\Sigma_{a\,full}^{tot}(E)$.

## 6. Conclusion

From the distribution (5) containing lifetime of statistical system (first-passage time, time before degeneration of system) we obtained the energy distributions of a system which contain the product of the Gibbs factor and a factor corresponding to the superstatistics which was introduced in [1] merely formally (though in [1] the dynamical justification is performed). In the superstatistics factor one averages not only over the values of the finite temperature, but over other system parameters as well, including the parameters of interaction with the environment. When passing from the initial distribution containing the lifetimes of a statistical system as an additional thermodynamic parameter, one used the assumptions about the shape of the system potential and the expression of the average lifetime from the stochastic storage model (Appendix A). A discrete version for discrete distribution is obtained, at which one changes argument $r_0 E \rightarrow (1-exp\{-r_0 E\})$ in, for example, Tsallis distributions. Below we consider the continuos distributions precisely generalizing the superstatistics, and show their correspondence to both known results and several novel trends of investigations.

The superstatistics theory [1,2] is the generalization of the nonextensive statistical mechanics [3, 4]. Some of the fields of applications of the nonextensive statistical mechanics are metastable states, Levy flights, systems of finite size [47]. It is shown in the present work that all these phenomena are conditioned physically by the finiteness of the lifetime, and by the possibilities of small probability big fluctuations which are generally dismissed when accounting for the system evolution.

The introduction of the lifetime (first-passage time) as a thermodynamic parameter is justified by the fact that finite systems possess finite lifetimes, which essentially influences their properties and the properties of their environment. The lifetime is interpreted as a fundamental quality having dual nature related to both the flux of external time, and the properties of a system. In [18] the expressions for the lifetime through dynamical characteristics of a system were obtained. These expressions show dependence on the interchange with the environment.

In [5, 6, 7, 8, 18] we pointed at the possibilities of describing nonequilibrium behaviour of arbitrary physical values (by which the system is open to the environment) with accounting for all factors influencing the interaction to the environment. The lifetime thermodynamics describes open systems far from equilibrium and can be applied to the investigation of the dissipative structures and other synergetic effects.

We managed to introduce a generalizing physical thermodynamical value – the lifetime of complex dynamical systems. The thermodynamic way of description aims at elucidating the general properties and statistical laws which do not depend on the peculiarities of the matter structure and are universal invariants. The contents of the thermodynamics are the set of such results. Since the finiteness of lifetime of arbitrary realistic systems is a universal property, we believe its inclusion into the thermodynamic description to be fruitful and necessary.

**Appendix A**

In [32] the behaviour of the random number of particles in a volume cell $\Delta V$ is modelled by a stochastic storage process [48] for the particle $Z(t)$ at the time $t$ in the volume $\Delta V$
$Z(t)=Z(0)+A(t) - \int_0^t r[Z(u)]du$, where $A(t)$ and $r[Z(t)]$ are arbitrary functions describing input and output from the system. The input value is described by the Levi process with non-decreasing trajectories and zero drift; at $A(0)\equiv 0$

$$E(exp\{-\theta A(t)\})=exp\{-t\varphi(\theta)\}\ ;\quad \varphi(\theta)=\int_0^\infty (1-exp\{-\theta x\})\nu(dx)\ ;\quad \nu(dx) = \lambda b(x)dx\ ;$$

$$\varphi(\theta)=\lambda-\lambda\psi(\theta);\quad \psi(\theta)=\int_0^\infty exp\{-\theta y\}b(y)dy;$$

$$\rho=\int_0^\infty x\nu(dx)=\int_0^\infty \lambda x b(x)dx=\lambda<x>\leq \infty\ ;\quad (A.1)$$

where $\lambda$ is the input intensity $b(x)$ is the distribution density of the input "batches", thus $\int_0^\infty \lambda x b(x)dx=<x>$ is average size of the input (per time unit). The lifetime is determined by the relation (2) and for the model with constant exit rate $a$ the average lifetime $<\Gamma(Z_0)>=Z_0/a(1-\rho/a)$, where $Z_0 = Z(t=0)$, $t_0$ is average time of an output from system of one particle. Assuming $Z_0$ to be some initial input. Lets average over it that is write using (A.1):

$$<\Gamma>=\int_0^\infty <\Gamma(Z_0)>b(Z_0)dZ_0=<x>/a(1-\rho/a)=\rho/\lambda a(1-\rho/a),\quad (A.2)$$

$$E(exp\{-s\Gamma_x\})=exp\{-x\eta(s)\};\ \eta(s)=s+\varphi(\eta(s));\ <\Gamma^2>_0-<\Gamma>^2_0=\rho\sigma^2/\lambda(1-\rho)^3;\ \sigma^2=\int_0^\infty x^2\lambda b(x)dx;$$

$$<Z>_{st}=\sigma^2 e/2(1-\rho)t_0,$$

where $t_0$ is average time of an output from system of one particle, $e$ is average energy of one particle of system. A statistical system possesses the state of the stationarity. The thermodynamical values at stationarity do not depend on time, and the average of the parameters does not change. Thereby $P_0=Q^{-1}=lim_{t\to\infty} P\{Z(t)=0|Z(0)=Z_0\}$, where $P_0$ is the probability of degeneration, $Q$ is grand canonical sum of the big canonical Gibbs stationary (close to equilibrium) ensemble. For the storage model with constant exit rate $P_0=0$, $\rho\geq a$; $P_0=1-\rho/a$, $\rho<a$.

Thus for the stable case with $\rho<a$, $\rho/a=1-P_0=1-Q^{-1}$. Inserting this value of $\rho/a$ into (A.2), we get: $<\Gamma>=\Gamma_0=(Q-1)/\lambda$, where $\Gamma_0$ means stationary lifetime value without perturbation. This expression coincides with (20).

The open system is considered, and the value $\rho$ represents some stochastic flow. If there are any stationary forces $v$, then $\varphi_f(y)=\varphi(y)-yf$, where $f$ is function of stationary force. Then $\rho_f=\rho-f$, and $-f=<I_v>$ is average value of the flux caused by action of force $f$. Thus

$$<\Gamma>=<x_f>/a_f(1-\rho_f/a_f) , \qquad (A.3)$$

where $<x_f>=\int xb_f(x)dx$. In a stationary case $dZ/dt=\rho_f-a_f(1-P_{0f})=0$, since $P_{0f}=1-\rho_f/a_f$. The value $\rho_f$ is equal to an average of the stationary flux in the system.

## Figure captions

Fig.1.Schematic time behaviour of a random process *y(t)* and lifetime *Γ*.

Fig.2. Schematic dependence of thermodynamic potential on order parameter $\eta$ in a case potential when the phase space of system is divided into isolated areas, each of which answers a metastable thermodynamic state.

Fig.3. Comparison $p(E)=E^{1/2}exp\{-25E\}+0,011exp\{-E/0,965\}sh(2,29E)^{1/2}$ and $m(E)=E^{1/2}exp\{-25E\}$; values of energy are specified in *eV*. Fig. 3(a): 0<E<0,5 eV; Fig. 3(b): 0<E<5 eV.

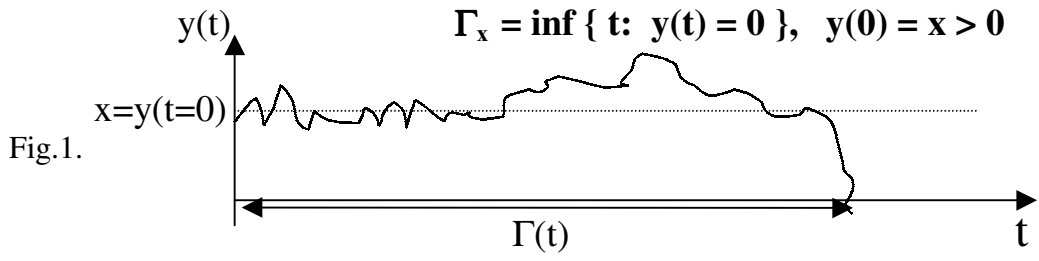

Fig.1.

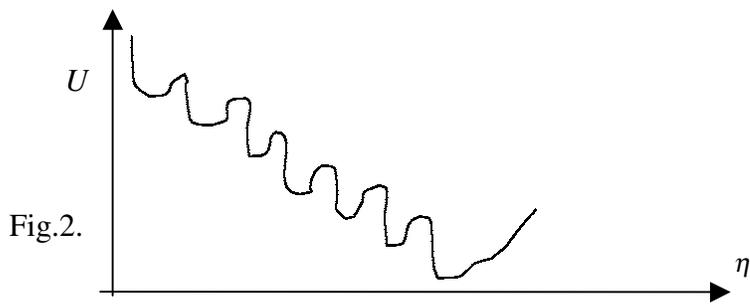

Fig.2.

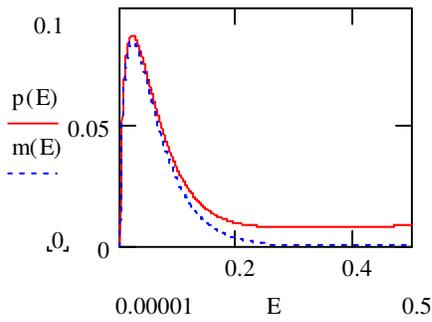 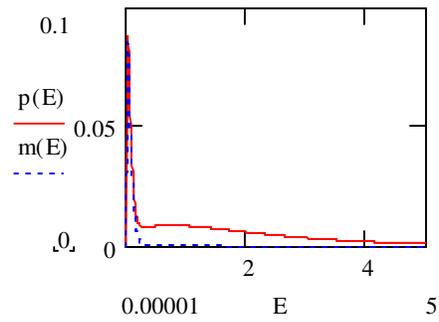

Fig.3 (a)  Fig.3 (b)